# Conceptual design of a tropical cyclone UAV based on the AR-6 Endeavor aircraft

**Chung-Kiak Poh[1], Chung-How Poh[2], Mei-Ling Yeh[1], and Tien-Yin Chou[1]**

[1] Geographic Information Systems Research Center, Feng Chia University, Taichung 40724, Taiwan

[2] Department of Physics, University of Newcastle, Callaghan, NSW 2308, Australia

Email address: kiak@gis.tw

## Abstract

This paper reports on the preliminary simulation work for a 1-meter class tropical cyclone UAV (unmanned aerial vehicle) based on the Formula-One AR-6 Endeavor full-sized airplane. Variants with different wing span are evaluated using the popular RealFlight® radio-control simulator. The 84 cm wing-span platform achieves a maximum cruising speed ($V_h$) of 407 $kmh^{-1}$ and demonstrates responsive flight controls throughout its flight envelope even in sustained wind speed of 225 $kmh^{-1}$. Being a small and agile UAV, it can be flown below the storm to measure surface wind directly and avoid possible uncertainty associated with the vertical wind profile adjustment. Chute-free vertical retardation technique is also proposed. The ultimate aim of the research is to develop cost effective UAVs that can be employed in a multi-agent setting to acquire high resolution data to enhance understanding of cyclogenesis and to make better predictions.

# Introduction

Tropical cyclones are intense atmospheric vortices that form over warm tropical oceans. It can be viewed as a dissipative structure that requires a constant input of energy in the form of latent heat. Tropical cyclones often cause widespread damage when they make landfall due to high winds, torrential rainfall and storm surges [1–3]. The main structural features of a tropical cyclone are the rainbands, the eye, and the eyewall [2,4]. The eye has a typical diameter of 32 to 64 km, though its formation mechanism is still not fully understood [4,5]. The eye is the calmest part of the storm with winds that usually do not exceed 24 kmh$^{-1}$ [4]. The eyewall region consists of a ring of tall thunderstorms that produce heavy rains and often the strongest winds [4].

In the vertical direction, winds are strongest near the surface and decay with height within the troposphere [6]. The super typhoon Haiyan that made landfall in Philippine in 2013 has sustained winds of 315 kmh$^{-1}$ with gusts as strong as 380 kmh$^{-1}$ [7]. Olivia which struck Australia in 1996 has the highest wind gust at 408 kmh$^{-1}$ though its highest 1-minute sustained wind was only 230 kmh$^{-1}$ [8]. Hurricane Katrina has a maximum sustained wind of 280 kmh$^{-1}$ [9]. The tops of a severe tropical storm can be over 15.2 km high [10].

The wind circulations of a matured tropical cyclone can be broadly divided into the primary and the secondary circulation [11]. The primary circulation refers to the tangential flow rotating about the central axis, and the secondary circulation refers to the "in-up-and-out circulation" (low and middle level inflow, upper-level outflow) [11]. Thus, the general air flow model of a tropical cyclone is air parcels spiraling inwards, upwards and outwards [11].

Manned aircrafts, such as the AT-6 Texan, B-17 Flying Fortress, WC-130 Hercules, and more recently the Lockheed WP-3D Orions and the Gulfstream IV SP Jet from NOAA, have been used to fly into and around tropical cyclones at a relatively safe height of 700 mbar level (approximately 3 km) to make measurements of interior barometric pressure and wind speed [12,13]. The data will enable forecasters to predict the strength of the storm, its trajectory and landfall location [13]. Of particular interest to forecasters and the public is the maximum sustained 10 m surface wind [14]. However, observations taken at elevated heights need to be properly adjusted in order to yield a reliable estimate of the surface winds based on the best knowledge of the vertical profile of wind in the core of the storm [14]. Aircrafts often fly at an angle into the crosswind (known as "crabbing") such that the resultant velocity vector is directed at the center of the storm [10]. Primary risk to manned flight is associated with downdrafts. Unusually strong downdrafts (~19 $ms^{-1}$) in the eyewall of Hurricane Emily (1987) were documented by in-situ aircraft measurements and a vertically pointing Doppler radar [15]. A near-mishap occurred in 1989 when a Lockheed WP-3D Orion was flown into Hurricane Hugo [16].

UAV approach has been investigated to reduce risks to flight crews and to explore new capabilities. The Aerosonde has a wing span of about 3 m and it made its first operational flight back in 1995 [17]. It penetrated the eyewall of the Typhoon Longwang on a reconnaissance observation mission in 2005 and the in situ wind measurement was consistent with the Doppler weather radar [18]. The recent Aerosonde Mark 4.7 has a dash speed of 150 $kmh^{-1}$ at sea level [19]. NASA begun employing the Global Hawk UAVs in 2007 to study tropical cyclone using the Airborne Vertical Atmospheric Profiling System (AVAPS), or more commonly known as the dropsondes, which were developed by the U.S National Center for Atmospheric Research (NCAR) [20]. The NASA Global Hawk is ideal as it is capable of flight altitudes greater than 16764 m and flight duration of up to 30 hours [20]. The expendable dropsondes is designed to be dropped from an aircraft at altitude to measure tropical cyclone conditions as it falls to the surface. From 50000 ft (15240 m) down to sea level, the descent rate varied from 29.46 to 11.68 $ms^{-1}$ (5800 to 2300 $ftmin^{-1}$). The dropsondes were available in two versions, the AVAPS II sonde and the mini sonde. The mini sonde measured 30.5 cm in length and 4.7 cm in diameter, with a mass of 165 g. The dropsondes cost approximately USD 350 each and about 1000 to 1500 of them are used annually [21,22].

In an attempt to make tropical storm UAVs more accessible and to reduce operating cost, we hereby propose an alternative approach and present the simulation work on a hurricane UAV design. The eventual prototype of the UAV is expected to be water-resistant, able to withstand the battering rain and wind within the eyewall. As we would like to have the airspeed comparable to the wind strength inside the eyewall, the UAV platform design was based on the speed record breaking Formula-One air racer, known as the AR-6 Endeavor which has a brilliant history including setting a new world record [23]. The proposed UAV was code-named Coriolis-7.

If successful, the future prototype of Coriolis-7 is intended to carry a suite of atmospheric sensors, similar to those of the dropsondes. A significant departure from the dropsondes is that, instead of using chutes, the Coriolis-7 will execute the stall-spin maneuver to achieve similar retardation in the vertical descent rate. The advantage is that the Coriolis-7 may exit the spin at any time, fly to another location and repeat the measurement sequence. In this way, the Coriolis-7 can acquire multiple vertical wind profile data in a single flight.

## Materials and Methods

### *Simulation details*

Simulation work was performed using the RealFlight® 6.5 simulator [24] running on a quad-core 2.2 GHz computer. It employs RealPhysics™ technology for accurate and realistic simulations [24]. The as-supplied simulation model AR-6 Endeavor (Fig. 1) was used as the base platform to create the Coriolis-7. The Endeavor model was powered by a 2-stroke 1.60 cu in glow engine, and it has a wing span of 2.23 m and flying weight of 4.96 kg. Modification of the Endeavor was done using the Accu-Model™ aircraft editor. In order to improve Coriolis-7 handling of the hurricane wind conditions, the wing span and chord were reduced so as to increase its wing-loading. Length of the horizontal stabilizer was shortened 32% to further reduce air drag. The landing gear was eliminated as well. Structural comparison between the original AR-6 and the Coriolis-7 is as shown in Fig. 2. Note that during simulation run, the original graphical model of AR-6 will be displayed despite physical properties, such as wing span, have been altered. The graphical model will need to be updated separately but for this work it was not done as it does not affect the flight physics central to this study. Table 1 summarizes the key specifications of both aircrafts. The primary components added to the platform to enable electric-powered flight were a brushless motor and a 12-cell 16 Ah lithium polymer (Li-po) battery pack. The key advantages of employing brushless electric motor are reliability in the presence of rain water, and insensitivity to change in air pressure. 3-axis angular position dependent (roll, pitch, yaw) gyro was added to the platform for flight stabilization. The completed model weighed in at 6.73 kg with a wing loading of 368.6 g dm$^{-2}$. Flight performance and stability was evaluated under different environmental conditions.

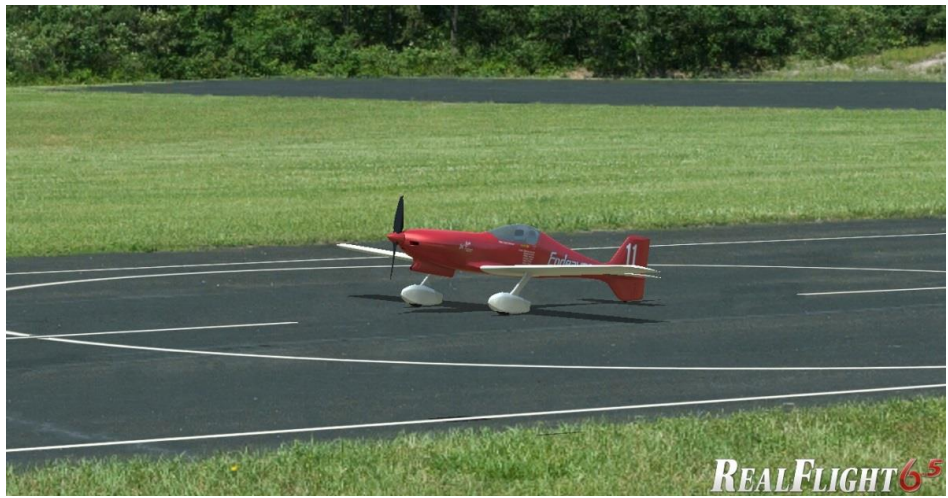

**Fig. 1** The AR-6 Endeavor model aircraft as supplied by the RealFlight simulator.

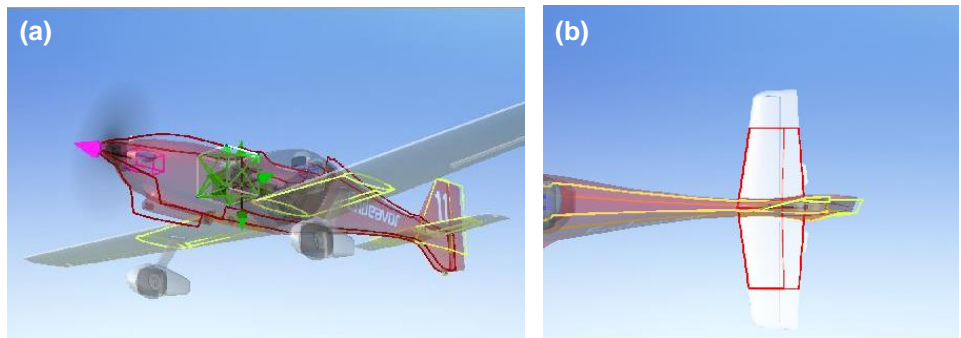

**Fig. 2** Comparison between the original AR-6 and the Coriolis-7 for the (a) wing section and (b) horizontal stabilizer.

**Table 1.** Specifications for the as-supplied AR-6 and the Coriolis-7

| Aircraft model | Flying weight (kg) | Fuselage width (m) | Wing span (m) | Wing loading (g dm$^{-2}$) |
|---|---|---|---|---|
| AR-6 | 4.96 | 0.251 | 2.23 | 89.1 |
| AR-6 electric | 7.90 | 0.130 | 2.23 | 142.4 |
| Coriolis-7 | 6.73 | 0.130 | 0.84 | 368.6 |

# Results and discussion

*Flight Simulation*

The electric version of the AR-6 has a $V_h$ (maximum speed in level flight with maximum continuous power) of 350 kmh$^{-1}$ while the Coriolis-7 achieved a $V_h$ of 407 kmh$^{-1}$, which is faster than the wind gust of the super Typhoon Haiyan (380 kmh$^{-1}$) but lower than the wind gust of the 1996 Olivia (408 kmh$^{-1}$). This, in principle, would allow the Coriolis-7 to be flown in tropical cyclones similar in strength to Typhoon Haiyan under the storm to enable direct measurement of the surface wind and other crucial parameters. The cruising speed and the corresponding electrical power of the AR-6 and Coriolis-7 are as shown in Fig. 3, with the last data points indicating their maximum airspeeds ($V_h$). The use of single propeller in the Coriolis-7 resulted in rolling tendency at full thrust when the launch velocity was less than 120 kmh$^{-1}$, such as in hand-launch mode, despite the best attempt of the onboard gyro to counteract. The use of the counter-rotating propellers reduced the undesirable propeller torque effect and the P-factor to a negligible level and there was no rolling tendency observed. This enhanced the flight stability throughout its flight envelope, and even allowed the UAV to be flown in the post-stall regime (< 85 kmh$^{-1}$) as shown in Fig. 3. However the $V_h$ of the platform was only 339 kmh$^{-1}$, probably due to higher propeller drag associated with co-axial propellers.

Given the importance of higher $V_h$ for better wind penetration, single propeller might be a better option. Afterall, the torque effect was not detrimental to the flight operation in regard to its hurricane missions. Furthermore, if the UAV were to be launched onboard a manned aircraft, the initial launch speed is likely to be higher. For example, the cruising speed of a Cessna 172 is 226 kmh$^{-1}$ [25] and at such speed simulation revealed that the propeller torque effect was insignificant, especially with a smooth and gradual ramping throttle profile.

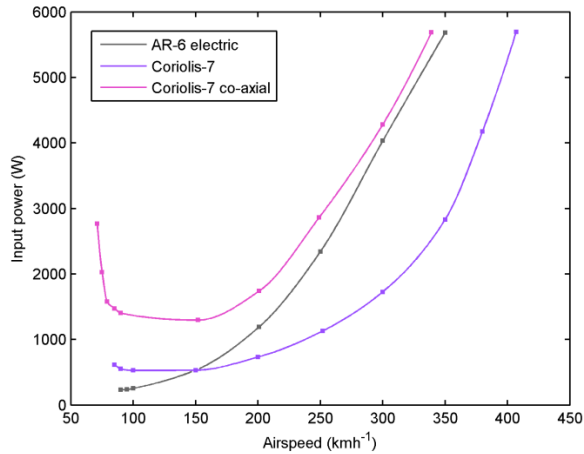

**Fig. 3** Plot of electrical power requirement as function of airspeed. The $V_h$ is represented by the final data point of each curve. Co-axial propellers enabled post-stall flight.

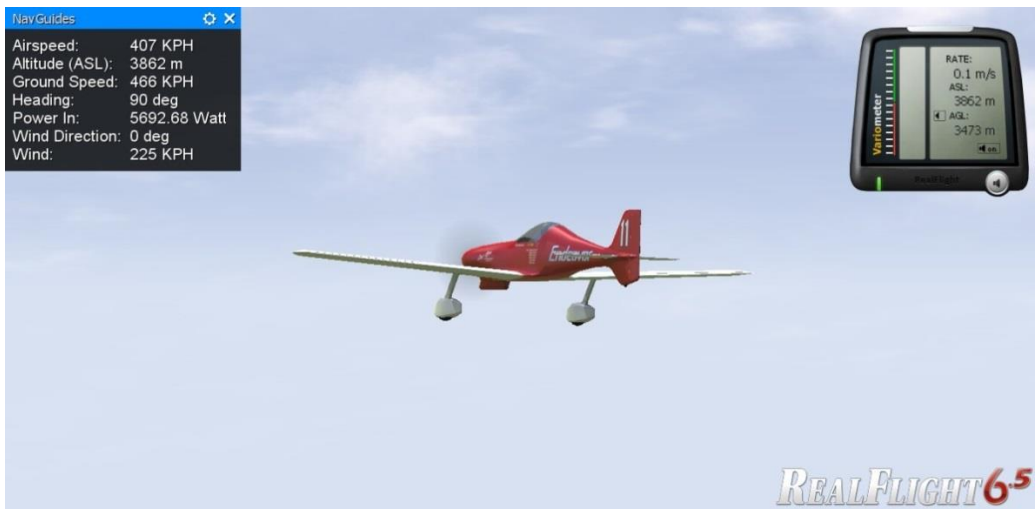

**Fig. 4** Crosswind test of Coriolis-7 in 225 kmh$^{-1}$ wind. The "NavGuides" include the direction of the wind, heading and airspeed of the airplane.

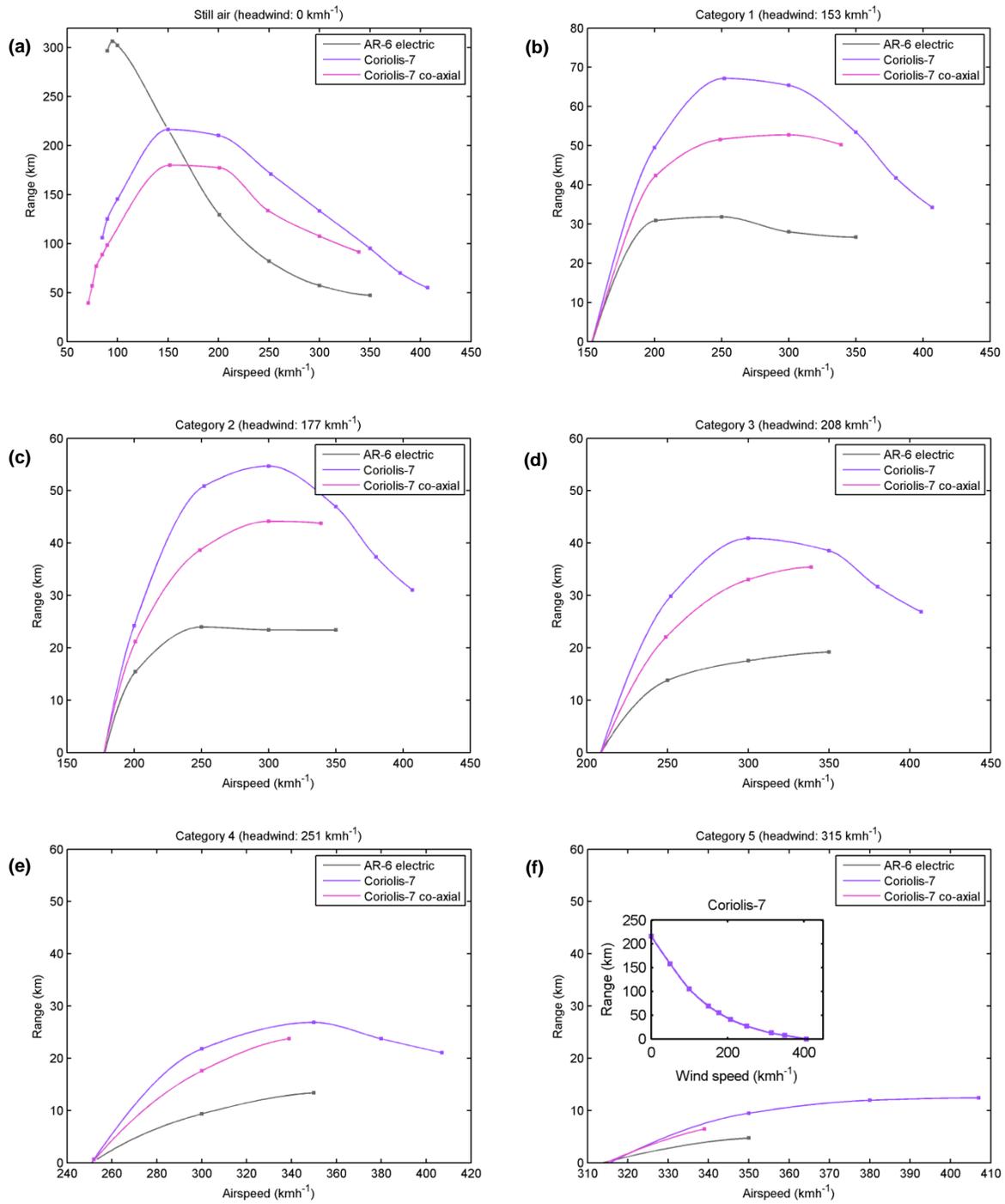

**Fig. 5** Range covered by the UAVs in different magnitude of headwind based on the Saffir-Simpson scale (a) 0 kmh$^{-1}$, (b) category 1, (c) category 2, (d) category 3, (e) category 4, and (f) category 5 in which the wind speed of Typhoon Haiyan was used. Inset in (f) shows the range covered by the Coriolis-7 at different headwind speeds.

The flight stability of the platforms under strong and gusty wind conditions was performed for different wind directions: headwind, tailwind, and crosswind. The platforms were subjected to horizontal wind speed of 225 kmh$^{-1}$, which is the maximum limit permitted by the simulator and with 100% turbulence and 30% wind gust. The control surfaces of the Coriolis-7 were found to be responsive under these conditions with no adverse flight characteristics. Whenever there were perturbations to the roll, pitch and yaw angles, the onboard 3-axis heading hold gyro would attempt to steer them back. Fig. 4 showed the Coriolis-7 flying at 407 kmh$^{-1}$ amidst a crosswind with speed of 225 kmh$^{-1}$. Note that ground speed of 466 kmh$^{-1}$ corresponds to the magnitude of the velocity vector, which is the vector sum of the velocity of the plane and the velocity of the wind. No inherent flight control instability was observed for both the AR-6 electric or the Coriolis-7, though the latter was more resistant to turbulence due to its higher wing loading.

Range of travel was derived from the flight power obtained in Fig. 3 and with battery capacity of 16 Ah. Fig. 5(a) shows the distance covered at different airspeed for the three platforms. The range covered was also studied in the presence of hurricane wind strength according to the Saffir-Simpson wind scale [26] in the worst case scenario (direct headwind). Fig. 5(b) to (f) show wind speeds for category 1 to 5, respectively, and the highest wind speed in each category was being used, with an exception of category 5. As category 5 is defined as having a wind speed of 252 kmh$^{-1}$ or more, it was thus represented by the sustained wind of the Super Typhoon Haiyan (315 kmh$^{-1}$).

In the absence of wind, the AR-6 covered the longest range of 306.6 km. Under hurricane wind conditions however, it was the Coriolis-7 that covered the furthest, followed by the Coriolis-7 with co-axial propellers. Interestingly, the most efficient cruising airspeed for the platforms was found to change with the intensity of headwind. For instance, the most efficient speed for the Coriolis-7 to fly in category 2 headwind was 300 kmh$^{-1}$ and it shifted to 350 kmh$^{-1}$ for category 4. The Coriolis-7 has a range of 12.4 km in category 5 headwind with the current battery setup. The inset in Fig 5(f) shows the range of Coriolis-7 in different headwinds.

While UAVs keep humans out of harm's way, the main limitation of battery-powered UAVs is the relatively short powered-flight duration. Lithium batteries typically have specific energy densities of only 0.4 to 0.9 MJkg$^{-1}$, versus 142 MJkg$^{-1}$ for liquid hydrogen [27,28]. Endurance UAVs powered by liquid hydrogen have repeatedly been demonstrated to achieve flight duration in range of days. Examples of hydrogen powered UAVs are the Global Observer, Ion Tiger and the Phantom Eye [29–31]. Gasoline has specific energy densities of approximately 46 MJkg$^{-1}$ and flight time of over 30 hours has been reported [17].

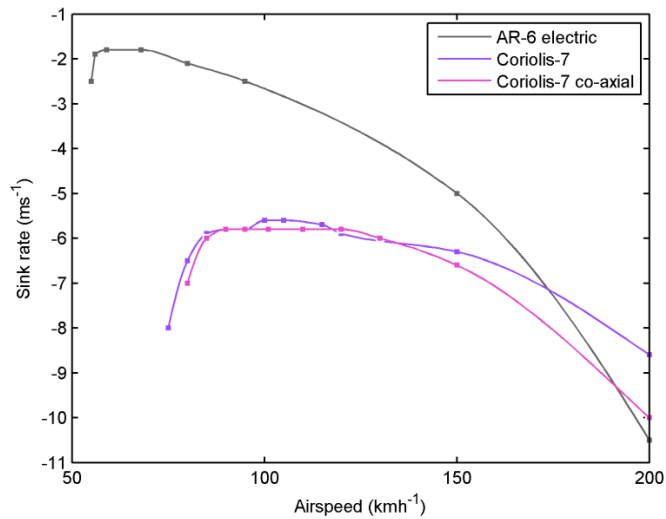

**Fig. 6** Glide polar curves for the model aircrafts

**Table 2.** Summary of gliding performance

| Aircraft model | Flying weight (kg) | Wing span (m) | Stall speed (kmh$^{-1}$) | Wing loading (g dm$^{-2}$) | Minimum sink rate (ms$^{-1}$) | Best glide speed (kmh$^{-1}$) | Maximum gliding ratio |
|---|---|---|---|---|---|---|---|
| AR-6 | 4.96 | 2.23 | 48.1 | 89.1 | 1.8 | 73 | 10.8:1 |
| Coriolis-7 | 6.73 | 0.84 | 85 | 368.6 | 5.6 | 166 | 6.8:1 |
| Coriolis-7 CA | 6.73 | 0.84 | 85 | 368.6 | 5.8 | 153 | 6.3:1 |

Fig. 6 shows the polar curves for the UAV platforms in 0 kmh$^{-1}$ wind. The minimum sink rate of the AR-6 and the Coriolis-7 are 1.8 ms$^{-1}$ and 5.6 ms$^{-1}$, respectively. The Coriolis-7 has much higher rate of sink because of its increased wing-loading. The maximum glide ratio and the best glide speed were derived from the gliding polar plot in Fig. 6. The gliding performance of the UAVs were summarized in Table 2. The Coriolis-7 with co-axial (CA) propeller has a slightly higher sinking rate (5.8 ms$^{-1}$) and stall speed than the single propeller system because of additional contribution of wind-milling drag from the second propeller. Both UAVs have stall speed around 85 kmh$^{-1}$.

*Chute-free vertical retardation technique*

Chutes were used in the dropsondes to slow their vertical descent rate and the descent rate of the dropsondes close to sea-level was 11.68 ms$^{-1}$. We would like to investigate a chute-free retardation technique. Eliminating the bulky chute will increase payload capacity and more importantly, enable the vertical descent sequence to be repeated multiple times per flight. The proposed method was based on an aerobatic maneuver known as the stall-spin. The UAV was deliberately entered into a stall to initiate the maneuver. The steady-state stall-spin was achieved by applying full deflection of rudder and aileron in the same direction and applying up elevator. The angular displacements of all three control surfaces were approximately 45°. The maneuver was performed under 500 m altitude. Fig. 7(a) shows a snapshot of the AR-6 electric performing the stall-spin with vertical descent rate of 10 ms$^{-1}$, which is even lower than that of the dropsondes (11.68 ms$^{-1}$). Having a significantly higher wing loading and narrower fuselage, the Coriolis-7 has a descent rate of 17.4 ms$^{-1}$. Optimization was performed to achieve a lower descent rate by increasing the wing span to 1.94 m, and the fuselage width was increased to 18 cm. The platform was denoted as Coriolis-7L. The resulting descent rate was 11.2 ms$^{-1}$, as indicated by the variometer as shown in Fig. 7(b). The Coriolis-7 CA has a descent rate of 15.8 ms$^{-1}$, assisted by the additional propeller drag. The stall-spin could be terminated by returning the control surfaces to neutral position and pulling up on the elevator to resume horizontal flight.

To investigate the ability of the technique to reliably detect wind speed, the stall-spin maneuver for the Coriolis-7L was repeated with various horizontal wind speeds. Turbulence and gust were turned off in order to evaluate the effect reliably. One could observe the platform drifting in the lateral direction with the wind. Fig. 8(a) and (b) showed the drifting in wind speed of 50 kmh$^{-1}$ and 180 km$^{-1}$, respectively. Ground speeds were found to closely match the wind speeds to within ±3 kmh$^{-1}$. This simulation work provided evidence that the stall-spin maneuver can be used as substitution for chute.

Videos showing the stall-spin maneuvers of the Coriolis-7L including the exit to level flight in still air and strong wind are available as ancillary files ([1] and [2], respectively). Also, it can be observed from the movie files that the axis of rotation of the stall-spin is close to the center of gravity of the aircraft and, therefore the maneuver was not expected to affect the empirical GPS speed measurement as the spatial variation is smaller than the resolution of GPS. The Coriolis-7L has a $V_h$ of 372 kmh$^{-1}$ which is intermediate between that of AR-6 electric and Coriolis-7. Summary of vertical descent rates and $V_h$ of the platforms are shown in Table 3.

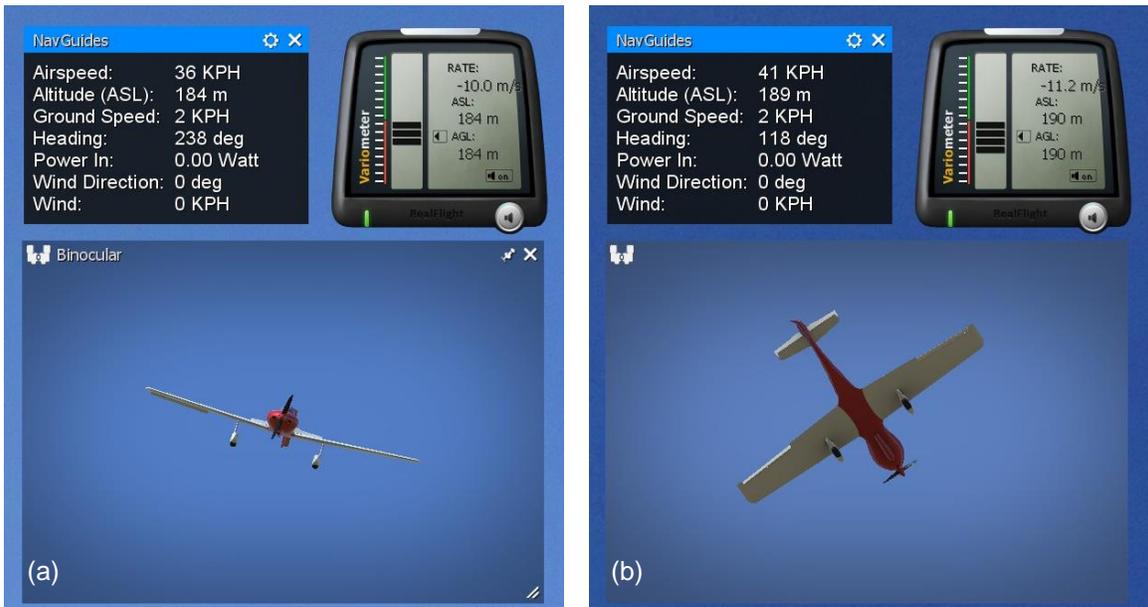

**Fig. 7** Dynamic parachuting: stall-spin maneuver performed by the (a) AR-6 electric and (b) Coriolis-7L with the variometers showing the corresponding descent rates. Full deflection of the control surfaces can be observed.

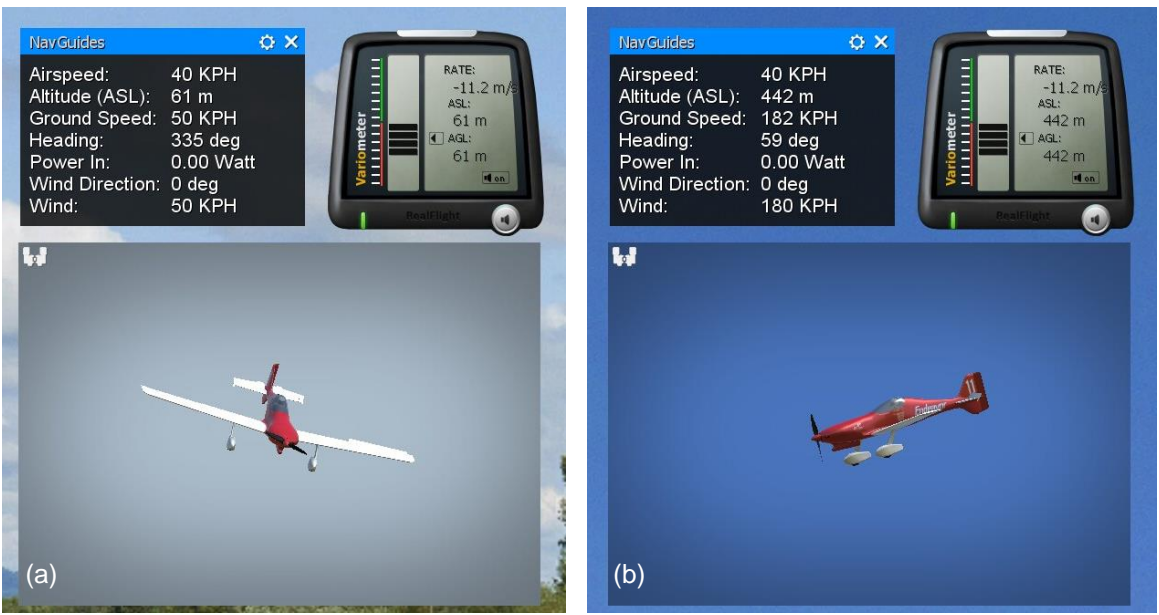

**Fig. 8** Coriolis-7 performing the stall-spin maneuver amidst wind of intensity a) 50 kmh$^{-1}$ and b) 180 kmh$^{-1}$. The ground speeds were found to correlate strongly with the wind intensities to within ±3 kmh$^{-1}$.

**Table 3.** Summary of $V_h$ values and stall-spin vertical descent rates for Coriolis-7 and its variants.

| Aircraft model | Wing span (m) | Fuselage width (cm) | Stall-spin descent rate [a] (ms$^{-1}$) | $V_h$ (kmh$^{-1}$) |
|---|---|---|---|---|
| AR-6 electric | 2.23 | 25.1 | 10.0 | 350 |
| Coriolis-7 | 0.84 | 13.0 | 17.4 | 407 |
| Coriolis-7 CA | 0.84 | 13.0 | 15.8 | 339 |
| Coriolis-7L | 1.94 | 18.0 | 11.2 | 372 |

[a] Descent rate of chute-operated dropsonde near sea-level is 11.68 ms$^{-1}$.

## Conclusions

We have proposed and simulated the conceptual design of a tropical cyclone UAV based on the AR-6 Formula-One air racer. In still air, the Coriolis-7 achieved a maximum cruising speed ($V_h$) of 407 kmh$^{-1}$. Flight controls were found to be responsive with wind of 225 kmh$^{-1}$ (upper limit of simulator) and with 30% wind gust. The proposed stall-spin maneuver was found to be an effective substitution for parachute with the advantage to perform multiple set of measurements per flight at different target locations compared to the dropsonde chute method.